\documentclass[lettersize,journal]{IEEEtran}
\usepackage{amsmath,amsfonts}
\usepackage[ruled]{algorithm2e}
\usepackage{array}
\usepackage[caption=false,font=normalsize,labelfont=sf,textfont=sf]{subfig}
\usepackage{textcomp}
\usepackage{stfloats}
\usepackage{url}
\usepackage{amssymb}
\usepackage{verbatim}
\usepackage{graphicx}
\usepackage{cuted}
\usepackage{bm} 
\usepackage{multirow}

\usepackage{cite}
\begin{document}

\title{\begin{huge} Error Performance of Coded AFDM Systems in Doubly Selective Channels\end{huge}}

\author{Haoran Yin
	\vspace{-2.5em}
	\thanks{
		This work was supported by Guangdong Natural Science Foundation
		under Grant 2019A1515011622. 
		
		 Haoran Yin is with the School of Electronics and Communication Engineering, Sun Yat-sen University, China (e-mail: yinhr6@mail2.sysu.edu.cn).

		}  
}
        % <-this % stops a space
\markboth{}%
{Shell \MakeLowercase{\textit{et al.}}: A Sample Article Using IEEEtran.cls for IEEE Journals}
\maketitle

\begin{abstract}
\textbf{Affine frequency division multiplexing (AFDM) is a strong candidate for the sixth-generation wireless network thanks to its strong resilience to delay-Doppler spreads. In this letter, we investigate the error performance of coded AFDM systems in doubly selective channels. We first study the conditional pairwise-error probability (PEP) of AFDM system and derive its conditional coding gain. Then, we show that there is a fundamental trade-off between the diversity gain and the coding gain of AFDM system, namely the coding gain declines with a descending speed with respect to the number of separable paths, while the diversity gain increases linearly. Moreover, we propose a near-optimal turbo decoder based on the sum-product algorithm for coded AFDM systems to improve its error performance. Simulation results verify our analyses and the effectiveness of the proposed turbo decoder, showing that AFDM outperforms orthogonal frequency division multiplexing (OFDM) and  orthogonal time frequency space (OTFS) in both coded and uncoded cases over high-mobility channels.}
\end{abstract}
\vspace{-0.3em}

\vspace{-1.0em}
\section{Introduction}
Emerging applications including space air-ground integrated networks (SAGIN), high-speed railways, and vehicle-to-vehicle (V2V) networks call for reliable communications techniques that adapt to high-dynamic scenarios. The widely adopted orthogonal frequency division multiplexing (OFDM) fails to support reliable communications in doubly selective channels due to the serious inter-carrier interference (ICI) induced by Doppler spreads \cite{23.11.14.1}, leading to an urgent demand for designing a new delay-Doppler-resilience waveform.

In this context, the recently proposed affine frequency division multiplexing (AFDM) attracts rapidly growing attentions \cite{23.10.17.1,bb101,23.10.17.2}. Along with some appealing features of inherent optimal diversity, low channel estimation overhead, and high backward compatibility with OFDM in doubly selective channels, AFDM is a promising candidate waveform for future wireless networks. Information symbols in AFDM are modulated on a set of orthogonal chirps via inverse \emph{discrete affine Fourier transform} (DAFT), where the chirps' slope are tuned elaborately to harvest an equivalent delay-Doppler (DD) channel representation in DAFT domain. Many researches have studied the implementation of practical AFDM systems concerning channel estimation \cite{23.10.17.2,23.10.17.4, 23.10.18.1}, signal detection \cite{bb102}, and multiple-input multiple-output techniques \cite{23.10.18.1,23.10.17.3}.

Channel coding is one of the most critical techniques to against fading and channel impairments and hence is widely-used in modern and pervious generations of wireless networks to ensure ultra-reliable communication. To the best of our knowledge, a comprehensive study on the performance of coded AFDM systems is still missing in the emerging AFDM literature. The main contributions of this letter are summarized as follows.
\begin{itemize} 
	\item[$\bullet$]
	We first study the conditional pairwise-error probability (PEP) of AFDM system and derive the corresponding conditional coding gain. Then we show how the number of resolvable paths, the Euclidean distance between the transmit codewords, and the maximum delay-Doppler spread of the channels influence the coding gain of coded AFDM systems with Monte Carlo simulation. In particular, we reveal that there is a fundamental trade-off between the diversity gain and the coding gain of AFDM system, i.e., the coding gain declines with a descending speed relative to the augmentation of resolvable paths, while the diversity gain increases linearly.
\end{itemize}
\begin{itemize} 
	\item[$\bullet$]
	To further improve the error performance of coded AFDM systems, we propose a near-optimal turbo decoder by exploring the sum-product \cite{23.11.15.1} algorithm. Based on that, the analytical results are verified  and a comparison among OFDM, orthogonal time frequency space (OTFS) \cite{23.10.16.3}, and AFDM is conducted, showing that AFDM exhibits the best performance in terms of frame error rate (FER) in both coded and uncoded cases over high-mobility channels.
\end{itemize}

\textit{Notations:} $\mathbb{C}$ denotes the set of complex numbers;
$a \sim \mathcal{C} \mathcal{N}\left(0, N_{0} \right)$ means that $a$ follows the complex Gaussian distribution with
zero mean and variance $N_{0}$; 
$\delta(\cdot)$ denotes the Dirac delta function; $\operatorname{diag}(\cdot)$ denotes a square diagonal matrix with the elements of input vector on the main diagonal;  $(\cdot)^{\mathrm{H}}$, $(\cdot)^{T}$, and $\left\| \cdot \right\|$ denote the Hermitian, transpose, and Euclidean norm operations; $\lvert \cdot \rvert$ denotes the absolute value of a
complex scalar; $(\cdot)_{N}$ denotes the modulus operation with respect to $N$;
 $Q(\cdot)$ denotes the tail distribution
function of the standard normal distribution.

\vspace{-1.0em}
\section{Coded AFDM System Model}
\label{sec2}
\begin{figure}[htbp]
	\vspace{-1.0em}
	\centering
	\includegraphics[width=0.397\textwidth,height=0.149\textwidth]{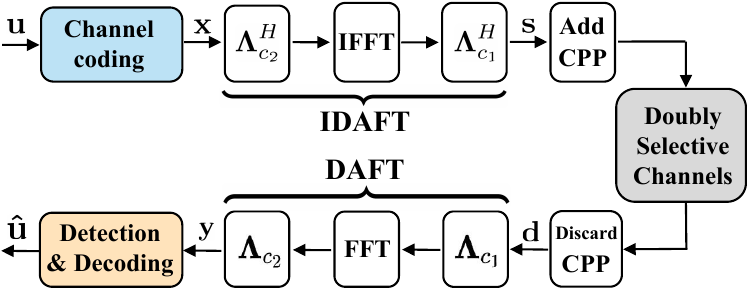}
	\vspace{-1.0em}
	\caption{The block diagram of coded AFDM system.}
	\vspace{-0.5em}
	\label{Fig2-1}
\end{figure}

In this section, we develop the coded AFDM system model with its block diagram shown in Fig. \ref{Fig2-1}.

\emph{\textbf{1) AFDM modulation}}: Let $T_{s}$ denotes sample period, $N$ denotes the number of subcarriers (chirps), then AFDM signal has a bandwidth $B=\frac{1}{T_{s}}$, subcarrier spacing $\Delta f=\frac{B}{N}=\frac{1}{NT_{s}}$. At the transmitter, an information sequence $\mathbf{u}$ is channel coded and mapped into DAFT domain vector $\mathbf{x} \in \mathbb{A}^{N \times 1}$, where $\mathbb{A}$ represents the modulation alphabet. Then, $N$-point inverse DAFT (IDAFT) is performed to modulate $\mathbf{x}$ to the time domain as \cite{23.10.17.2}
\vspace{-0.3em}
\begin{equation}
	\vspace{-0.3em}
	\vspace{-0.2em}
	s[n]= \sum_{m=0}^{N-1} x[m] \phi_{m}[n],  \ n=0, \cdots, N-1
	\label{eq2-1}
\end{equation}
where $n$ and $m$ denote the time and DAFT domains indices, respectively, chirp subcarrier $\phi_{m}[n]$ is given by $\phi_{m}[n]=\frac{1}{\sqrt{N}} e^{j 2 \pi\left(c_{1} n^{2}+c_{2} m^{2}+  n m / N\right)}$,
$c_{1}$ and $c_{2}$ are two AFDM parameters, and $c_{1}$ determines the chirps' slope. Equation (\ref{eq2-1}) can be written in matrix form as
\vspace{-0.3em}
\begin{equation}
	\vspace{-0.3em}
	\mathbf{s} =  \boldsymbol{\Lambda}_{c_{1}}^{H} \mathbf{F}^{H} \boldsymbol{\Lambda}_{c_{2}}^{H} \mathbf{x} = \mathbf{A}^{H} \mathbf{x}
	\label{eq2-2}
\end{equation}
where $\mathbf{A} = \boldsymbol{\Lambda}_{c_{2}} \mathbf{F} \boldsymbol{\Lambda}_{c_{1}}\in\mathbb{C}^{N\times N}$ represents the DAFT matrix,  $\mathbf{F}$ is the DFT matrix with entries $e^{-j 2 \pi m n / N} / \sqrt{N}$, $\boldsymbol{\Lambda}_{c}\triangleq\operatorname{diag}\left(e^{-j 2 \pi c n^{2}}, n=0,1, \ldots,   N-1\right)$. Before transmitting $\mathbf{s}$, an \emph{chirp-periodic} prefix (CPP) should be added, which plays the same role as the cyclic prefix (CP) in OFDM to cope with the multipath propagation and makes the channel lie in a periodic domain equivalently.

\emph{\textbf{2) Channel model}}:
Consider the general doubly selective channel with following  impulse response  with delay $\tau$ and Doppler $\kappa$ as 
	$h(\tau, \kappa)=\sum_{i=1}^{P} h_{i} \delta\left(\tau-l_{i}T_{s}\right) \delta\left(\kappa-\nu_{i}\Delta f\right)$ \cite{23.10.16.3}, 
where $P$ is the number of paths, $h_{i}$ denotes the channel coefficient of the $i$-th path, non-negative integer
$l_{i} \in [0, l_{\max}]$ is the associated delay
normalized with $T_{s}$,  $\nu_{i}=\alpha_{i}+\beta_{i}$ represents the associated Doppler shift normalized with subcarrier spacing $\Delta f$ and has a finite support bounded by $[-\nu_{\max}, \nu_{\max}]$, $\alpha_{i} \in [-\alpha_{\max}, \alpha_{\max}]$ and $\beta_{i} \in (-\frac{1}{2}, \frac{1}{2}]$ are the integer and fractional parts of $\nu_{i}$ respectively, $\nu_{\max}$ denotes the maximum Doppler and $\alpha_{\max}$ denotes its integer component.

\emph{\textbf{3) AFDM demodulation}}:
At the receiver,  the relationship between the received time domain symbols $\mathbf{d}$ and $\mathbf{s}$ can be expressed as $
	d[n]=\sum_{i=1}^{P} h_{i} e^{-j  \frac{2 \pi}{N}\nu_{i}n} s[(n-l_{i})_{ N}]+v[n]$, 
where  $v \sim \mathcal{C} \mathcal{N}\left(0, N_{0}\right)$ represents the additive white gaussian noise (AWGN) component. It can be vectorized as
\vspace{-0.5em}
\begin{equation}
	\vspace{-0.3em}
	\mathbf{d} = \sum_{i=1}^{P} h_{i} \mathbf{\tilde{H}}_{i}  \mathbf{s}  + \mathbf{v}= \mathbf{\tilde{H}} \mathbf{s} + \mathbf{v}  
	\label{eq2-6}
\end{equation}
where $\mathbf{v} \sim \mathcal{C} \mathcal{N}\left(\mathbf{0}, N_{0} \mathbf{I}_{N}\right)$ is the time domain noise vector, $\mathbf{\tilde{H}}\in\mathbb{C}^{N\times N}$ denotes the effective time domain channel matrix, $\mathbf{\tilde{H}}_{i}= \boldsymbol{\Delta}_{\nu_{i}} \boldsymbol{\Pi}^{l_{i}}$ represents the time domain subchannel matrix of the $i$-th path (each path can be viewed as one subchannel), $\boldsymbol{\Pi}$ denotes the forward cyclic-shift matrix which models the delay, while the digital frequency shift matrix $\boldsymbol{\Delta}_{\nu_{i}} \triangleq \operatorname{diag}\left(e^{-j \frac{2 \pi}{N} \nu_{i} n}, n=0,1, \cdots, N-1\right)$ models the Doppler. Finally, $N$-point DAFT is implemented and  $\mathbf{d}$ are transformed to the DAFT domain symbols $\mathbf{y}$ with $
	y[m]=\sum_{m=0}^{N-1} d[n] \phi_{m}^{*}[n]   + w[m]$,
where $w$ represents the noise in the DAFT domain. Its matrix representation  is 
\vspace{-0.5em}
\begin{equation}
	\vspace{-0.3em}
	\mathbf{y} = \boldsymbol{\Lambda}_{c_{2}} \mathbf{F} \boldsymbol{\Lambda}_{c_{1}} \mathbf{d} = \mathbf{A} \mathbf{d}.
	\label{eq2-8}
\end{equation}
Since $\mathbf{A}$ is a unitary matrix, $w$ has the same statistical properties as $v$.

\emph{\textbf{4) Input-output relationship}}:
The matrix form of AFDM input-output relationship in the DAFT domain can be obtained by substituting (\ref{eq2-2}) and  (\ref{eq2-6}) into (\ref{eq2-8}) as \cite{23.10.17.2}
\vspace{-0.3em}
\begin{equation}
	\vspace{-0.3em}
	\label{eq2-12-2}
	\mathbf{y} =\mathbf{H}_{\text{eff}}\mathbf{x} + \mathbf{w}= \sum_{i=1}^{P} h_{i} \mathbf{H}_{i} \mathbf{x} + \mathbf{w} 
\end{equation}
where $\mathbf{H}_{i} = \mathbf{A} \mathbf{\tilde{H}}_{i} \mathbf{A}^{H}$ denotes the \textbf{DAFT domain subchannel matrix} of the $i$-th path, $\mathbf{H}_{\text{eff}} = \sum_{i=1}^{P} h_{i} \mathbf{H}_{i}$ is the effective channel matrix, $\mathbf{w}\sim \mathcal{C} \mathcal{N}\left(\mathbf{0}, N_{0} \mathbf{I}_{N}\right)$ is the DAFT domain noise vector. 

\emph{\textbf{Remark 1:}} It has been proven in \cite{23.10.17.2} (Theorem 1) that AFDM can achieve optimal diversity in doubly selective channels as long as $c_{1} = \frac{2 (\alpha_{\max }+k_{\nu})+1}{2N}$, and $c_{2}$ is set as an arbitrary irrational number (spacing factor $k_{\nu}$ is a non-negative integer used to combat the fractional Doppler).

\vspace{-0.7em}
\section{Error Analysis of Coded AFDM Systems}
\vspace{-0.3em}
\label{sec3}
In this section, we investigate the theoretical error performance of coded AFDM systems. For the convenience of illustration, we denote the vectors of channel coefficient, delay indices and Doppler indices as $\mathbf{h}$, $\bm{\tau}$, and $\bm{\kappa}$, respectively, i.e., $\mathbf{h} = [h_{1}, h_{2}, \ldots, h_{P}]^{T} \in \mathbb{C}^{P\times 1}$, $\bm{\tau} = [l_{1}, l_{2},\ldots, l_{p}]^{T}\in \mathbb{C}^{P\times 1}$, and $\bm{\kappa} = [\nu_{1}, \nu_{2},\ldots, \nu_{p}]^{T}\in \mathbb{C}^{P\times 1}$. Perfect channel state information (CSI) and maximum likelihood (ML) detection are assumed at the receiver. Then, according to \cite{23.10.17.2}, Equation (\ref{eq2-12-2}) can be presented in an alternate way as 
\vspace{-0.3em}
\begin{equation}
	\vspace{-0.3em}
	\mathbf{y} = \boldsymbol{\Phi}_{\bm{\tau},\bm{\kappa}}(\mathbf{x}) \mathbf{h}+\mathbf{w}
	\label{eq23-11-16-1}
\end{equation}
where $
\mathbf{\Phi}_{\bm{\tau},\bm{\kappa}}(\mathbf{x}) = \left[\mathbf{H}_{1} \mathbf{x}, \ \mathbf{H}_{2} \mathbf{x}, \ \ldots, \  \mathbf{H}_{P} \mathbf{x} \ \right]\in \mathbb{C}^{N \times P}$
is the equivalent codeword matrix. 
Therefore, for a given channel realization, the   \textbf{conditional PEP} of transmitting the symbol $\mathbf{x}$ and deciding in favor of $\mathbf{x^{\prime}}$ at the
receiver can be expressed as \cite{23.11.16.1}
\vspace{-0.3em}
\begin{equation}
	\vspace{-0.3em}
	P\left(\mathbf{x}, \mathbf{x}^{\prime} \mid \mathbf{h},\bm{\tau},\bm{\kappa} \right)=Q\left(\sqrt{\frac{\left\|
				\mathbf{\Phi}_{\bm{\tau},\bm{\kappa}}(\mathbf{e})\mathbf{h}\right\|^{2}}{2 N_{0}}}\right)
		\label{eq23-11-16-3}
\end{equation}
where $\mathbf{e} = \mathbf{x}-\mathbf{x}^{\prime}$ is the corresponding \textbf{codeword difference sequence}. Define \textbf{codeword difference matrix} $\boldsymbol{\Omega}_{\bm{\tau},\bm{\kappa}}(\mathbf{e}) \triangleq \mathbf{\Phi}_{\bm{\tau},\bm{\kappa}}(\mathbf{e})^{H}\mathbf{\Phi}_{\bm{\tau},\bm{\kappa}}(\mathbf{e})$, which is a Hermitian matrix and hence can be  diagonalized by unitary transformation as $\boldsymbol{\Omega}_{\bm{\tau},\bm{\kappa}}(\mathbf{e}) = \mathbf{U}^{H}\mathbf{\Lambda}\mathbf{U}$, where $\mathbf{\Lambda}=\text{diag}\{\lambda_{1}, \lambda_{2}, \ldots, \lambda_{P}\}$,  $\lambda_{i}$ is the $i$-th nonnegative real eigenvalue (sorted in the descending order) of $\boldsymbol{\Omega}_{\bm{\tau},\bm{\kappa}}(\mathbf{e})$. Then, we have
\vspace{-0.3em}
\begin{equation}
	\vspace{-0.3em}
	\begin{aligned}
		\left\|
		\mathbf{\Phi}_{\bm{\tau},\bm{\kappa}}(\mathbf{e})\mathbf{h}\right\|^{2}
		&= \mathbf{h}^{H}\mathbf{\Phi}_{\bm{\tau},\bm{\kappa}}(\mathbf{e})^{H}\mathbf{\Phi}_{\bm{\tau},\bm{\kappa}}(\mathbf{e})\mathbf{h} \\
		&= \mathbf{h}^{H}\mathbf{U}^{H}\mathbf{\Lambda}\mathbf{U}\mathbf{h}
		=\mathbf{\tilde{h}}^{H} \mathbf{\Lambda}\mathbf{\tilde{h}}
		= \sum_{i=1}^{r} \lambda_{i}|\tilde{h}_{i}|^{2}
	\end{aligned}
	\label{eq23-11-16-5}
\end{equation}
with $\mathbf{\tilde{h}} = \mathbf{U}\mathbf{h}$, $\tilde{h}_{i}$ being the $i$-th value of $\mathbf{\tilde{h}}$, $r$ denoting the rank of $\boldsymbol{\Omega}_{\bm{\tau},\bm{\kappa}}(\mathbf{e})$. Substituting (\ref{eq23-11-16-5}) into (\ref{eq23-11-16-3}) and applying the Chernoff bound of the Q-function, i.e., $Q(\gamma) \leq \exp \left(-\frac{1}{2} \gamma^{2}\right), \forall \gamma>0$, we have
\vspace{-0.3em}
\begin{equation}
	\vspace{-0.3em}
	\begin{aligned}
		P\left(\mathbf{x}, \mathbf{x}^{\prime} \mid \mathbf{h},\bm{\tau},\bm{\kappa} \right) &\leq \text{exp}(-\frac{\sum_{i=1}^{r} \lambda_{i}|\tilde{h}_{i}|^{2}}{4N_{0}})
	\end{aligned}
	\label{eq23-11-16-4}
\end{equation}
where $\frac{1}{N_{0}}$ denotes the signal-to-noise ratio (SNR). Moreover, applying the inequality of $\exp(-\gamma)\leq \frac{1}{1+\gamma}, \ \forall \gamma \geq 0$, (\ref{eq23-11-16-4}) can be  expanded as
\vspace{-0.5em}
\begin{equation}
	\vspace{-0.5em}
	\begin{aligned}
		P\left(\mathbf{x}, \mathbf{x}^{\prime} \mid \mathbf{h},\bm{\tau},\bm{\kappa} \right) 
		\leq
		\prod_{i=1}^{r}\frac{1}{1+\frac{\lambda_{i}|\tilde{h}_{i}|^{2}}{4N_{0}}}.
	\end{aligned}
	\label{eq23-11-16-6}
\end{equation}
Considering that $h_{i}$ follows the distribution of $\mathcal{C N}(0,1 / P)$ (uniform scattering profile and $|h_{i}|$ follows the Rayleigh distribution) and $\mathbf{U}$ is unitary, $\tilde{h}_{i}$ follows the distribution of $\mathcal{C N}(0,1 / P)$ as well. Therefore, in the case of \textbf{Rayleigh fading} and high SNR, (\ref{eq23-11-16-6}) can be further simplified as
\vspace{-0.3em}
\begin{equation}
	\vspace{-0.3em}
	\begin{aligned}
		P\left(\mathbf{x}, \mathbf{x}^{\prime} \mid \bm{\tau},\bm{\kappa} \right) 
		\leq
		\frac{1}{(\frac{1}{4N_{0}})^{r}\prod\limits_{i=1}^{r}\frac{\lambda_{i}}{P}}.
	\end{aligned}
	\label{eq23-11-16-7}
\end{equation}
It should be noted that (\ref{eq23-11-16-7}) is consistent with the analysis in \cite{23.10.17.2}.
Generally, the power of SNR $\frac{1}{N_{0}}$ is defined as the \textbf{diversity gain}, which
dominates the exponential behaviour of the error performance for AFDM systems with respect to SNR. According to \textbf{Remark 1}, AFDM systems always attain optimal diversity gain, i.e., $r=P$ regardless of ($\bm{\tau}$, $\bm{\kappa}$). Therefore, after some manipulations on (\ref{eq23-11-16-7}), we have
\vspace{-0.3em}
\begin{equation}
	\vspace{-0.3em}
	\begin{aligned}
		P\left(\mathbf{x}, \mathbf{x}^{\prime} \mid \bm{\tau},\bm{\kappa} \right) 
		\leq
		\left(\frac{1}{4N_{0}}\right)^{-P}\left(\frac{\left(\prod_{i=1}^{P}\lambda_{i}\right)^\frac{1}{P}}{P} \right)^{-P}.
	\end{aligned}
	\label{eq23-11-16-8}
\end{equation}
In particular, we define the term $\frac{\left(\prod_{i=1}^{P}\lambda_{i}\right)^\frac{1}{P}}{P}$ as \textbf{conditional coding gain}, which indicates the potential error performance improvement introduced by channel coding for a given channel realization. It is determined by the number of paths $P$, delay-Doppler profile ($\bm{\tau}$, $\bm{\kappa}$) and codeword difference sequence $\mathbf{e}$ jointly. 

\begin{figure}[tbp]
	\centering
	\includegraphics[width=0.4\textwidth,height=0.332\textwidth]{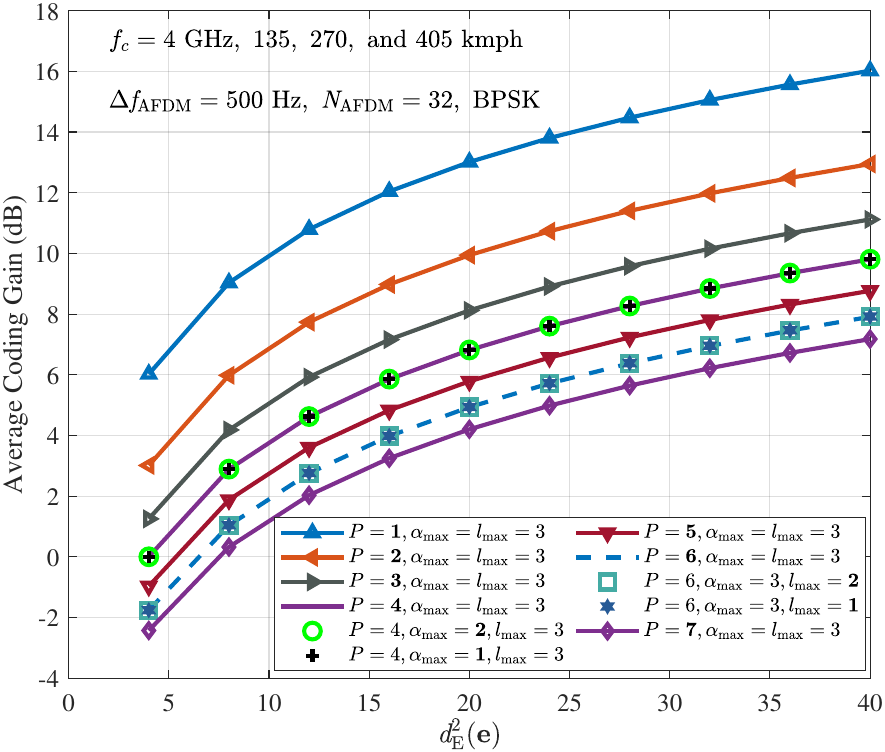}
	\vspace{-1.0em}
	\caption{Average coding gain versus $d_{\text{E}}^{2}(\mathbf{e})$ with different number of paths, maximum delay and Doppler.}
	\vspace{-1.9em}
	\label{Fig3-1}
\end{figure}

In order to derive the unconditional coding gain, one should find the statistical distribution of the term $\prod_{i=1}^{P}\lambda_{i}$ regarding ($\bm{\tau}$, $\bm{\kappa}$) and $\mathbf{e}$, which is generally intractable \cite{23.11.16.2}. Therefore, we resort to the Monte Carlo method to take a deep look at the unconditional coding gain of AFDM systems by approximating it with averaged conditional coding gain. Fig. \ref{Fig3-1} shows the \textbf{average coding gain} versus \textbf{squared error Euclidean distance} $d_{\text{E}}^{2}(\mathbf{e})=\mathbf{e}^{H}\mathbf{e}$ with different  maximum delay and Doppler\footnote{Without loss of generality, we consider BPSK mapping and generate $\mathbf{e}$  by randomly selecting the indices of $\mathbf{e}$ and set them to “$\pm2$", where the number of non-zero indices is  $d_{\text{E}}^{2}(\mathbf{e})/4$, and $\mathbf{e}$ is of the form $[0,2,0,-2,0,\ldots,0]^{T}$. Besides, the integer delay and Doppler indices are chosen randomly according to the uniform distribution among $[0, l_{\max}]$ and $[-\alpha_{\max}, \alpha_{\max}]$, respectively.}. 
We can observe that with the rise of $d_{\text{E}}^{2}(\mathbf{e})$, the average coding gain increases with a descending speed. Moreover, we can notice that different maximum Doppler $\alpha_{\max}$ and maximum delay $l_{\max}$ do not have a prominent influence on the average coding gain. This implies that given the same number of paths, the error performance of AFDM systems remain nearly unchanged with various channel dynamic levels (the maximum speeds corresponding to $\alpha_{\max}=1$, $2$, and $3$ are 135 kmph, 270 kmph, and 405 kmph, respectively). More importantly, we can observe clearly that with the increases of $P$,  the average coding gain decreases with a descending speed. This indicates that there
exists a fundamental trade-off between the coding gain and the diversity gain, which are framed formally as follows.

\emph{\textbf{Corollary 1 (Trade-Off Between Coding Gain and Diversity Gain of AFDM Systems)}:} For a given channel code, the coding gain of AFDM systems declines at a descending speed as the number of separable paths increases, while the diversity gain grows linearly.

\textbf{Corollary 1} implies that when there are few notable propagation paths in the channel, the diversity gain of AFDM systems is small and the potential of error performance improvement offered by channel coding is of great tremendousness. While in rich-scattering conditions, it is expected that channel coding can only provide a limited improvement to the overall error performance of AFDM systems.

\vspace{-0.8em}
\section{Turbo Decoder for Coded AFDM Systems}
\vspace{-0.3em}
\label{sec4}
To unleash the potential of channel coding to improve the error performance of AFDM systems to the greatest extent, we propose a near-optimal turbo decoder by exploring the sum-product algorithm (SPA) \cite{23.11.15.1} for coded AFDM systems.

\emph{\textbf{1) Symbol-wise SPA detector:}} We first derive the symbol-wise SPA detection for AFDM systems. For the convenience of illustration, we consider integer delay and Doppler ($\beta_{i}=0, \forall i \in [1,P]$). In this case,  due to the delay-Doppler spread and fading of the doubly selective channel, each received symbol $y[m^{\prime}]$ consists of $P$ impaired transmitted symbols $x[m_{i}], m_{i} = m^{\prime}+\operatorname{ind}_{i}$,  $i \in [1,P]$, where $
	\operatorname{loc}_{i} \triangleq\left(\alpha_{i}+2 N c_{1} l_{i}\right)_{N}$
is the \textbf{index indicator} of the $i$-th path \cite{23.10.17.2}.
Meanwhile, each transmitted symbol affects $P$ received symbols. Denote the set of $P$ received symbols that influenced by $x[m]$ as $\mathbf{y}_{m}\triangleq\{y\left[\left(m-	\operatorname{loc}_{i}\right)_{N}\right] \mid i \in [1,P]\}$, and the set of $P-1$ transmitted symbols that related to the received symbol $\mathbf{y}_{m}[i]$ except for $x[m]$ as $\mathbf{x}_{m}^{(i)} \triangleq \{x\left[\left(m-\operatorname{loc}_{i}+\operatorname{loc}_{j} \right)_{N} \right] \mid j \in [1,P], j\neq i \}$. Consider symbol-wise \textit{maximum a posterior (MAP)} criterion, i.e., 
\vspace{-0.7em}
\begin{equation}
	\vspace{-0.7em}
	\hat{x}[m]=\arg \max_{x[m] \in \mathbb{A}} \operatorname{Pr}\{x[m] \mid \mathbf{y}\}
	\label{eq23-11-17-1}
\end{equation}
where the a posterior probability  can be expended by applying the Bayes's rule as
\vspace{-0.7em}
\begin{equation}
	\vspace{-0.7em}
\operatorname{Pr}\{x[m] \mid \mathbf{y}\}
 \propto 
 \operatorname{Pr}\{\mathbf{y} \mid x[m]\} \operatorname{Pr}\{x[m]\}. 
	\label{eq23-11-17-3}
\end{equation}
Let $\mathbf{y}_{m}|_{i+1}^{P}$ represents the vector of the $(i+1)$-th to the $P$-th entries of $\mathbf{y}_{m}$, then (\ref{eq23-11-17-3}) can be further expended with the chain rule as
\begin{equation}
	\begin{aligned}
&\operatorname{Pr}\{\mathbf{y} \mid x[m]\} \operatorname{Pr}\{x[m]\} \\ 
&=\prod_{i=1}^{P} \operatorname{Pr}\left\{\mathbf{y}_{m}[i]  \Big| \mathbf{y}_{m} |_{i+1}^{P}, \left.\mathbf{y} \backslash \mathbf{y}_{m}, x[m]\right\}\right\} \operatorname{Pr}\{x[m]\} \\
&=\prod_{i=1}^{P} \sum_{\mathbf{x}_{m}^{(i)}} \operatorname{Pr}\left\{\mathbf{y}_{m}[i], \mathbf{x}_{m}^{(i)} \Big| \mathbf{y}_{m}|_{i+1}^{P}, \mathbf{y} \backslash \mathbf{y}_{m}, x[m]\right\} \operatorname{Pr}\{x[m]\}
\end{aligned}
\label{eq23-11-17-4}
\end{equation}
where $\mathbf{y} \backslash \mathbf{y}_{m}$ represents the complementary set of $\mathbf{y}_{m}$ correspending to $\mathbf{y}$. Applying the chain rule again, we have
\vspace{-0.5em}
\begin{equation}
	\vspace{-0.7em}
	\begin{aligned}
&\operatorname{Pr}\{\mathbf{y} \mid x[m]\} \operatorname{Pr}\{x[m]\} \\
&=\prod_{i=1}^{P} \sum_{\mathbf{x}_{m}^{(i)}} \operatorname{Pr}\left\{\mathbf{y}_{m}[i] \Big| \mathbf{x}_{m}^{(i)}, x[m]\right\} \\ 
& \qquad \qquad \  \times \operatorname{Pr}\left\{\mathbf{x}_{m}^{(i)} \Big| \mathbf{y}_{m}|_{i+1}^{P}, \mathbf{y} \backslash \mathbf{y}_{m}\right\} \operatorname{Pr}\{x[m]\} .
	\end{aligned}
	\label{eq23-11-17-5}
\end{equation}
Finally, assuming that the entries of $\mathbf{x}_{m}^{(i)}$ are independent to the
elements from $\mathbf{y}_{m}|_{1}^{i-1}$, we obtain\footnote{According to \cite{23.11.15.1} and \cite{23.11.17.1}, this assumption will only introduce few errors when the channel is sufficiently sparse ($N\gg P $).}
\vspace{-0.5em}
\begin{equation}
	\vspace{-0.6em}
	\begin{aligned}
		&\operatorname{Pr}\{x[m] \mid \mathbf{y}\} 
		\approx 
		\prod_{i=1}^{P} \sum_{\mathbf{x}_{m}^{(i)}} \operatorname{Pr}\left\{\mathbf{y}_{m}[i] \Big| \mathbf{x}_{m}^{(i)}, x[m]\right\} \\ 
		& \qquad \qquad \  \  \quad \qquad \qquad \times \operatorname{Pr}\left\{\mathbf{x}_{m}^{(i)} \Big| \mathbf{y} \backslash \mathbf{y}_{m}\right\} \operatorname{Pr}\{x[m]\}
	\end{aligned}
	\label{eq23-11-17-6}
\end{equation}
where 
\vspace{-0.5em}
\begin{equation}
	\vspace{-0.4em}
	\operatorname{Pr}\left\{x[m] \mid \mathbf{y} \backslash \mathbf{y}_{m}\right\} \propto \prod_{\substack{j=1 \\ j \neq i}}^{P} \operatorname{Pr}\left\{x[m] \mid \mathbf{y}_{m}[j]\right\}	
	\label{eq23-11-17-8}
\end{equation}
\begin{equation}
	\vspace{-0.3em}
	\begin{aligned}
		\operatorname{Pr}\{\mathbf{y}_{m}[i]    \mid  \mathbf{x}_{m}^{(i)},& x[m]\}=\frac{1}{\sqrt{\pi N_{0}}} \\
		&\times\exp \left(-\left|\mathbf{y}_{m}[i]-\mathcal{F}(m,i)\right|^{2} \Big/ N_{0}\right)
	\end{aligned}
	\label{eq23-11-17-7}
\end{equation}
\vspace{-0.5em}
\begin{equation}
	\vspace{-0.5em}
	\begin{aligned}
\mathcal{F}(m,i) = &\sum_{j=1}^{P-1} \mathbf{H}_{\text{eff}}\left[\textit{ind}_{\mathbf{y}}(m,i),\textit{ind}_{\mathbf{x}}(m,i,j)\right] \mathbf{x}_{m}^{(i)}[j]\\
&+\mathbf{H}_{\text{eff}}\left[\textit{ind}_{\mathbf{y}}(m,i),m\right] x[m]
\end{aligned}
\end{equation}
two index extractors are defined as $\textit{ind}_{\mathbf{y}}(m,i) = \left(m-\operatorname{ind}_{i}\right)_{N}$ and $\textit{ind}_{\mathbf{x}}(m,i,j)=\left(m-\operatorname{loc}_{i}+\operatorname{loc}_{j} \right)_{N}$. The detailed procedures of the symbol-wise SPA detector are summarized in Algorithm \ref{algorithm-1}.

\begin{figure}[htbp]
	\centering
	\includegraphics[width=0.49\textwidth,height=0.12\textwidth]{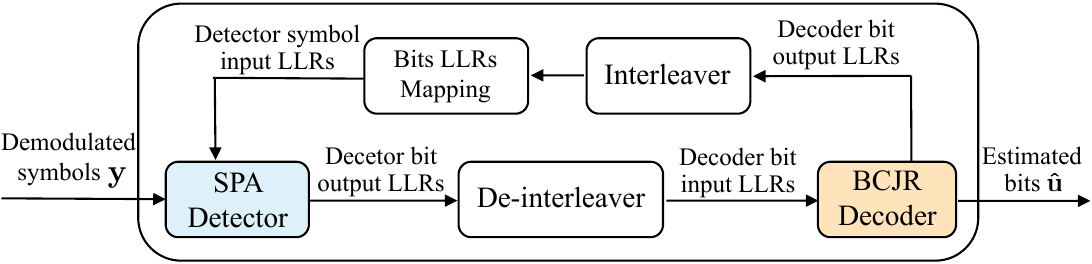}
	\vspace{-2.0em}
	\caption{The proposed iterative turbo decoder for coded AFDM systems.}
	\label{Fig4-1}
	\vspace{-1.0em}
\end{figure}

\emph{\textbf{2) Near-optimal turbo decoder:}} Based on the above SPA detector, we propose an iterative turbo decoder for coded AFDM systems, which is illustrated in Fig. \ref{Fig4-1}. After acquiring the a posterior probability of  $\hat{x}[m], m=0, \cdots, N-1$, we can calculate the corresponding bit log likelihood ratios (LLR) with $\mathbb{A}$, which is de-interleaved and then fed to an optimal 
decoder, e.g., Bahl-Cocke-Jelinek-Raviv (BCJR) decoder \cite{23.11.15.2} for convolutional code. The output bit LLRs from the BCJR decoder are then interleaved and converted to symbol LLRs according to $\mathbb{A}$, which is the updated a priori probability 
$\operatorname{Pr}\{x[m]\}$ and initiates the SPA detector again. This completes one turbo iteration.

Since both the symbol-wise SPA detector and the BCJR decoder follow  the MAP criterion, the proposed turbo decoder is expected to approach the optimal error performance of AFDM systems in terms of the bit error ratio (BER). Moreover, the BCJR decoder therein can be replaced by other good decoders depending on the adopted channel code, e.g.,  low-density parity-check (LDPC) and turbo codes.

\begin{algorithm}
	\caption{Symbol-wise SPA detector for AFDM}
	\label{algorithm-1}

	\LinesNumbered  % 开始显示行号
	\KwIn{$\mathbf{y}$, $\mathbb{A}$, $N$, $P$, $\mathbf{H}_{\text{eff}}$, $\bm{\tau}$, $\bm{\kappa}$, maximum number of iteration $I_{\max}$, \textit{a prior probability} $\operatorname{Pr}\{x[m]\}$.}
	\For{$I = 1,2,\cdots, I_{\max}$}{
		\For{$i = 1,2,\cdots, P$}{
			\For{$m = 0,1,\cdots, N-1$}{
				\textbf{Step 1:}  Find out all combinations of $\mathbf{x}_{m}^{(i)}$ with respect to $\mathbb{A}$.\\
				
				\textbf{Step 2:} For each possible combination of $\mathbf{x}_{m}^{(i)}$ calculate (\ref{eq23-11-17-7}) and $\operatorname{Pr}\left\{\mathbf{x}_{m}^{(i)} \Big| \mathbf{y} \backslash \mathbf{y}_{m}\right\}$ with (\ref{eq23-11-17-8}).\\
				
				\textbf{Step 3:} calculate $\operatorname{Pr}\{x[m] \mid \mathbf{y}\}$ with (\ref{eq23-11-17-6}).\\
			}
		}
	}
	Make hard decision of $\hat{x}[m], m=0, \cdots, N-1$ according to (\ref{eq23-11-17-1}).\\
	\KwOut{$\mathbf{\hat{x}}$ and $\operatorname{Pr}\{x[m] \mid \mathbf{y}\}$.}
\end{algorithm}

\vspace{-0.7em}
\section{Simulation Results}
\vspace{-0.3em}
\label{sec5}
In this section, we present the error performance of coded AFDM in terms of FER. Carrier frequency $f_{c}=4$ GHz, number of subcarriers $N=128$, subcarrier spacing $\Delta f_{\text{AFDM}}=$ 500 Hz, BPSK and 4QAM mappings, and the proposed SPA detector and turbo decoder are applied. Without loss of generality, we adopt three near-$1/2$ coderate convolutional codes, termed “Code A”, “Code B”, and “Code C”, with generating polynomials given by (3,1), (5,7), and (51,77) and corresponding minimum squared Euclidean distances $d_{\text{min, E}}^{2}(\mathbf{e})$ given by $12$, $20$, and $32$,  respectively. We only consider integer delay and Doppler with Rayleigh fading, where $l_{\max}=3$ and $\alpha_{\max}=3$,  corresponds to a maximum UE speed of 405 kmph. For each channel realization, the channel coefficient $h_{i}$ follows the distribution of $\mathcal{C N}(0,1 / P)$ and the delay and Doppler indices are chosen randomly according to the uniform distribution among $[0, l_{\max}]$ and $[-\alpha_{\max}, \alpha_{\max}]$, respectively.

\begin{figure*}[htbp]
	\vspace{-1.0em}
	\centering
	\includegraphics[width=1\textwidth,height=0.36\textwidth]{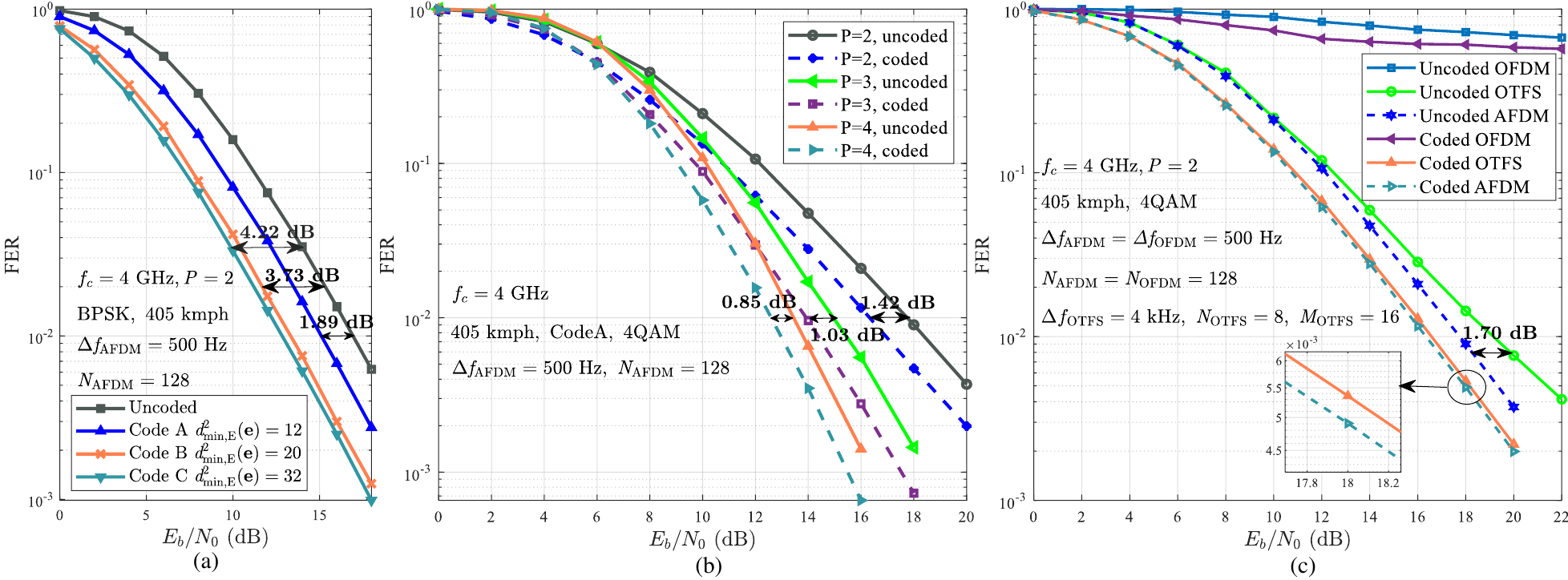}
	\vspace{-2.5em}
	\caption{(a) FER performance of AFDM systems with different codes with different codes, (b) FER performance of AFDM systems with different number of paths, and (c) FER comparison among OFDM, OTFS, and AFDM in uncoded and coded cases.}
	\label{Fig5-1}
	\vspace{-1.5em}
\end{figure*}
Fig. \ref{Fig5-1} (a) shows the FER performance of AFDM systems with different codes. $P=2$ and the result of uncoded AFDM is also provided. We can observe that the FER curves of uncoded AFDM system and coded AFDM systems with different minimum squared Euclidean distances share the same diversity slope. This is because the uncoded AFDM already achieves the optimal diversity gain and applying channel coding does not influence the overall diversity gain of AFDM systems, as revealed in (\ref{eq23-11-16-8}). Moreover, we can also notice that with the  increase of $d_{\text{min, E}}^{2}(\mathbf{e})$, the channel gain of coded AFDM system enhances.  Therefore, a preliminary guideline for the code design of AFDM systems is to maximize $d_{\text{min, E}}^{2}(\mathbf{e})$ among all pairs of codewords of the adopted channel code. Furthermore, the trend of coding gain enhancement over larger $d_{\text{min, E}}^{2}(\mathbf{e})$ gets slow down with the increase of $d_{\text{min, E}}^{2}(\mathbf{e})$. These observations are consistent with our analyses of Fig .\ref{Fig3-1}.

Fig. \ref{Fig5-1} (b) shows the FER performance of AFDM systems with different number of paths. 4QAM and code A are applied. We can see that the diversity gains of the uncoded and coded cases are the same regardless of the value of $P$. Moreover, the diversity gain increases with the increase of $P$, which demonstrates the effectiveness of the symbol-wise SPA detector and turbo decoder in exploring the optimal diversity of AFDM. More importantly, the coding gain of AFDM system declines at a descending rate. In specific, at FER $=10^{-2}$,  the coding gains of AFDM system with $P=2$, $3$, and $4$ are 1.43 dB, 1.03 dB, and 0.83 dB, respectively. This verifies the correctness of \textbf{Corollary 1}.

We next compare the error performance of OFDM, OTFS, and AFDM with the same time-frequency resources. Frequency domain single-tap SPA detection is used in OFDM systems and we can see from Fig. \ref{Fig5-1} (c) that the OFDM systems are paralyzed due to the serious ICI. Moreover, We can notice that the uncoded OTFS can not achieve optimal diversity, which is consistent with the analyses in \cite{23.11.18.1}. This leads to a huge error performance gap between uncoded OTFS and uncoded AFDM. In specific, at FER $\approx 8 \times 10^{-3}$, the required SNR of uncoded OTFS is around 1.70 dB larger than that of uncoded AFDM. By observing the FER trends of the two systems, we can infer that the error gap between uncoded OTFS and AFDM system will become larger with the increase of $E_{b}/N_{0}$. Furthermore, we can see that the difference between coded OTFS and coded AFDM is smaller than the uncoded cases, which is consistent with the conclusion in \cite{23.11.18.2}  that the achieved diversity gain of OTFS systems can be improved via channel coding. It is worth noting that the coded AFDM still exhibits slightly better error performance than the coded OTFS. Therefore, we can conclude that AFDM outperforms OFDM and OTFS in both uncoded and coded cases.

\vspace{-1.0em}
\section{Conclusion}
\label{sec6}
In this letter, we provide a comprehensive study on the error performance of coded AFDM systems in doubly selective channels. We derive the conditional coding gain of AFDM systems and show that there is a  fundamental trade-off between coding gain and diversity gain in AFDM systems. Moreover, we explore the sum-product algorithm and design a near-optimal turbo decoder for AFDM systems. Simulation results verify our analyses and the effectiveness of the proposed SPA detector and turbo decoder. Finally, a comparison among OFDM, OTFS, and AFDM is conducted, showing that AFDM outperforms OTFS in both uncoded and coded systems, especially in the former case. In the near future, we will investigate the modern code design for AFDM systems with imperfect channel state information.

\vfill
\end{document}